\documentclass{IEEEtran}%[9pt,technote]
\usepackage{amsmath}
\usepackage{graphicx, amssymb}
\usepackage[dvips]{epsfig}
\usepackage{color}
\newtheorem{asu}{Assumption}
\newtheorem{theorem}{Theorem}
\newtheorem{remark}{Remark}

\newcommand{\carrew} {\hfill $\Box$}
\newcommand{\carre}{\begin{flushright} \rule{2mm}{2mm} \end{flushright}}
    \title{ Bounded Inputs Total Energy Shaping for Mechanical Systems   }
\author{
M.~Reza~J.~Harandi,  Amir Molaei, Hamid D. Taghirad and Jose Guadalupe Romero 

\thanks{ \textsuperscript{a}\textbf{A}dvanced            \textbf{R}obotics and \textbf{A}utomated \textbf{S}ystems (ARAS), Faculty of Electrical Engineering,
	K. N. Toosi University of Technology,
	P.O. Box 16315-1355, Tehran, lran; {\tt\small jafari@email.kntu.ac.ir, taghirad@kntu.ac.ir}
}  
\thanks{	\textsuperscript{b} Concordia University, Montral, Canada; {\tt\small a\_molaei@encs.concordia.ca} }
\thanks{ \textsuperscript{c} Departamento Acad\'emico de Sistemas Digitales,ITAM, R\'io Hondo 1,  01080, Ciudad de M\'exico, M\'exico {\tt\small jose.romerovelazquez@itam.mx} }
}

\begin{document}
		\maketitle
	\thispagestyle{empty}
	\pagestyle{empty}

    %\author{
    %    \name{M.~Reza~J.~Harandi\textsuperscript{a}\thanks{CONTACT M.~Reza~J.~Harandi. Email:jafari@email.kntu.ac.ir}
    %           , Hamid D. Taghirad\textsuperscript{a}\thanks{CONTACT Hamid D. Taghirad. Email:aghirad@kntu.ac.ir} and Jose Romero\textsuperscript{b} \thanks{CONTACT Jose Romero. Email:should be defined} }
    %       \affil{\textsuperscript{a} \textbf{A}dvanced
    %                  \textbf{R}obotics and \textbf{A}utomated \textbf{S}ystems (ARAS),
    %                   Faculty of Electrical Engineering,
    %                   K. N. Toosi University of Technology,
    %                  P.O. Box 16315-1355, Tehran, lran; \textsuperscript{b} should be defined   }
    %     }

    \begin{abstract}
Designing control systems with bounded input is a practical
consideration since realizable physical systems are limited by the
saturation of actuators. The actuators' saturation degrades the
performance of the control system, and in extreme cases, the
stability of the closed-loop system may be lost. However, actuator
saturation is typically neglected in the design of control systems,
with compensation being made in the form of over-designing the
actuator or by post-analyzing the resulting system to ensure
acceptable performance. The bounded input control of fully actuated
systems has been investigated in multiple studies, but it is not
generalized for under actuated mechanical systems. This article
proposes a systematic framework for finding the upper bound of
control effort in underactuated systems, based on interconnection
and the damping assignment passivity based control (IDA-PBC)
approach. The proposed method also offers design variables for the
control law to be tuned, considering the actuator's limit. The major
difficulty in finding the control input upper bounds is the velocity
dependent kinetic energy related terms. Thus, the upper bound of
velocity is computed using a suitable Lyapunov candidate as a
function of closed-loop system parameters. The validity and
application of the proposed method are investigated in detail
through two benchmark systems.

    \end{abstract}
%\begin{IEEEkeywords}
%        underactuated mechanical systems; bounded input; total energy
%        shaping
%\end{IEEEkeywords}

\section{Introduction}
Underactuated mechanical systems (UMS) are characterized by the fact
that they have fewer actuators than their degrees of freedom (DOF).
In other words, the size of the configuration space exceeds the size
of the control input space~\cite{he2019underactuated}. Examples of
such systems include, but not limited to, legged robots, swimming
robots, flying robots, continuum robots, and gantry cranes as well
as manipulator with structural flexiblity. Underactuated robots
(URs) are expected to be more energy-efficient and lightweight than
fully actuated counterparts while still provide the adequate
dexterity without reducing the reachable configuration
space~\cite{xu2008sliding}. While UMS have broad applications in
different disciplines,  due to the reduced control inputs, the
traditional non-linear control methods are not directly applicable.
The control of UMS has been investigated through several control
methods including feedback linearization, sliding mode,
backstepping, fuzzy control, and passivity-based control (PBC).
Except for the PBC approach, none of the mentioned methods have
succeeded to be generalized to the UMS.

The term PBC was initially introduced in \cite{ortega1989adaptive}
as a control approach to make the closed-loop system passive. The
PBC of UMS was first presented in \cite{ortega2002interconnection}
as the extension of the energy-shaping plus damping injection
technique of \cite{takegaki1981new}, which was proposed to solve
state feedback set point regulation problems in fully actuated
robotic systems. The essence of passivity-based control is energy
shaping \cite{ortega2000energy}, which provides a natural procedure
to shape the potential energy in the closed-loop Euler--Lagrange
(EL) systems \cite{ortega2002stabilization}. Energy shaping control
treats the plant and its controller as energy-transformation
devices. In the trajectory tracking of the mechanical systems, the
energy shaping control based on the classical procedure of PBC does
not provide an EL representation of the closed-loop system, in other
words, the total energy of the closed-loop is not a physical energy
equation. Additionally, it cannot be used for stabilization of UMS
as it is not suitable for total energy shaping and is mostly used
for potential energy shaping. Nevertheless, an extended approach
introduced in \cite{ortega2002stabilization} as Interconnection and
damping assignment passivity based control (IDA-PCB) offers a
physically inspired generalized method for the control of UMS, in
terms of solvability of two PDEs that correspond to the potential
and kinetic energy shaping. By using this method, the dynamics of a
mechanical system is expressed in the form of the Port Hamiltonian
(PH) model with desired interconnection and damping matrices and a
Hamiltonian function, which is the summation of desired potential
and kinetic energy.

One of the most challenging aspects of the IDA-PBC is the analytical
solution of the PDEs, which have been studied in various studies
\cite{borja2015shaping,harandi2020matching,romero2016energy}. Moreover, the
bounded input stabilization of UMS in the presence of input
saturation is another challenging but importany issue that is rarely
addressed. The concept of bounded input control applies to a wide
variety of control applications. It is due to the fact that control
inputs of realizable systems are constrained by the saturation of
the actuator. Saturation of the actuator degrades the performance of
the control systems and can lead to instability in the closed
loop~\cite{zergeroglu2000adaptive}. A common drawback seen in many
of the control methods is the large magnitudes for the input torque,
which, in fact, is not taken into account in the design procedure of
the controller~\cite{spong1986control}. Actuator saturation is
typically neglected in the design of control systems, with
compensation being made in the form of over-designing of the
actuator or by post-analyzing the resulting system to ensure
acceptable performance.

In contrast to fully actuated robots, the bounded input
stabilization of URs is less attended by the researchers. To the
best of the authors' knowledge, \cite{santibanez2005control} is the
only reported paper in the field of IDA-PBC approach, which is used
for an inertia wheel pendulum that has bounded inputs. In this
article, the system has a constant mass matrix, leading to the total
energy being manipulated by a constant inertia matrix. By this means
all terms related to kinetic energy shaping are cancelled. However,
generally the mass matrix is configuration dependant and the kinetic
energy terms appears in the energy shaping formulation. As the
kinetic energy is quadratic with respect to velocity, the upper
bound of energy cannot be simply calculated if the upper bound of
the velocity is not known. In this article we present a generalized
methodology to calculate the upper bound of the control input in UMS
based on IDA-PBC. The proposed framework also provides design
parameters to be tuned to restrict the control effort within the
actuator saturation. To this aim, the upper bounds of the velocity
terms in the control law are derived to find the maximum kinetic
energy of the closed-loop system.

The IDA-PCB control law requires solving two PDEs, which in general
have numerous solutions. In some conditions, the solution of PDEs
which appears in the control law is non-smooth. Considering this,
finding the upper bound of the IDA-PBC control law with the proposed
framework allows using a suitable solution of the PDEs for the
formulation of the control law, in order to keep the control effort
within an expected limit. This application of the proposed method is
further investigated through an example for which two solutions for
the corresponding PDEs are found. Additionally we propose a
two-phase control methodology for non-smooth control laws, in which
a primary controller brings the state of the system into a region
far from the singular points, and then the secondary controller is
designed and implemented based on IDA-PBC.

The remaining structure of this paper is as follows; in Section
\ref{s2}, we present an overview of the IDA-PBC for the ease of
reference. Next, in Section \ref{s3}, we discuss a methodology to
find the upper bound of the control input based on IDA-PBC approach.
This is done by the analysis of every single terms of the control
law, the potential energy related terms and kinetic energy related
terms.  Finally, in Section \ref{s4} the applicability of the method
is investigated in two separate case studies.

\textbf{Notation.} Unless determined, the vectors in this paper are
column vector including the gradient of functions. For an arbitrary
matrix     $\xi\in\mathbb{R}^{n\times m}$ and vector
$a\in\mathbb{R}^n$, $\xi_i$ and $a_i$ denotes the $i$th row and
$i$th element of $\xi$
    and $a$, respectively,  $\xi_{(i,j)}$ is the $(i,j)$th element of
    that matrix, $I_n$ denotes an $n\times n$ identity matrix and
    $0_{m\times n}$ is an $m\times n$ matrix of zeros. Furthermore, $\mathbb{R}^+$ and
    $\mathbb{R}^-$ denote the set of positive and negative real numbers, and $\lambda_m\{\xi\}$
    and $\lambda_M\{\xi\}$ denote the maximum and minimum eigenvalues of
    the square matrix $\xi$, respectively.

\section{IDA-PBC for Mechanical Systems with Physical Damping}\label{s2}
For the ease of reference, a brief review  of IDA-PBC for mechanical
systems in the
    presence of natural damping (frictional forces) based on \cite{gomez2004physical} is
    given here. The dynamic equations of a mechanical system with physical damping in
    the port Hamiltonian representation is as follow:
    \begin{equation}
    \label{1}
    \begin{bmatrix}
    \dot{q} \\ \dot{p}
    \end{bmatrix}
    =\begin{bmatrix}
    0_{n\times n} & I_n \\ -I_n & -R(q)
    \end{bmatrix}
    \begin{bmatrix}
    \nabla_q H \\ \nabla_p H
    \end{bmatrix}
    +\begin{bmatrix}
    0_{n\times m} \\ G(q)
    \end{bmatrix}
    \tau
    \end{equation}
    in which $n$ is the DOF and $m$ is the number of control inputs, $q,p\in \mathbb{R}^n$ are generalized position and momentum,
    respectively, and $M\dot{q}=p$, in which $M^T(q)=M(q)>0$ is the inertia matrix. Moreover, $H(q,p)=\frac{1}{2}p^TM^{-1}(q)p+V(q)$ is the total energy
    of the mechanical system which is the summation of the kinetic and potential energy, $G(q)\in \mathbb{R}^{n\times m}$ is the input coupling
    matrix and
    $R(q)\in \mathbb{R}^{n\times n}$ is the smooth positive semi-definite damping matrix.
    In IDA-PBC it is assumed that the desired structure of the closed-loop system is as follows:
    \begin{equation}
    \label{2}
    \begin{bmatrix}
    \dot{q} \\ \dot{p}
    \end{bmatrix}
    =
    \begin{bmatrix}
    0_{n\times n} & M^{-1}M_d \\ -M_dM^{-1} & J_2-R_2
    \end{bmatrix}
    \begin{bmatrix}
    \nabla_q H_d \\ \nabla_p H_d
    \end{bmatrix}
    \end{equation}
    where $M_d$ and $H_d$ are the desired mass matrix and the desired total energy of the closed loop system, respectively. Additionally $J_2(q,p)$ is a skew-symmetric matrix and $R_2$ is a positive semi-difinite matrix in the following form:
    \begin{align}
    R_2=\frac{1}{2}(RM^{-1}M_d+M_dM^{-1}R)+GK_vG^T.
    \label{4.5}
    \end{align}
    the desired equilibrium point $q^*$ should also satisfy:
    \begin{equation}
    \label{1.5}
    q^*=\text{arg min} V_d(q).
    \end{equation}
    Notice that in the rest of the paper, we assume that $V_d$ is designed such that at the desired equilibrium point $(q^*)$, $H_d=0$.
    The control law proposed in the IDA-PBC method is defined as \eqref{3} which is obtained by setting \eqref{1} and \eqref{2} equal.
    \begin{align}
    \label{3}
    \tau=(G^TG)^{-1}G^T\bigg(\nabla_q V-M_dM^{-1}\nabla_q V_d+\nabla_q K&\nonumber\\-M_dM^{-1}\nabla_q K_d+(J_2-GK_vG^T)\nabla_p H_d\bigg)&
    \end{align}
 The above mentioned control law requires $M_d$ and $V_d$ which are to be obtained through analytical solution of \eqref{4a} and \eqref{4b}, respectively, given as below:
\begin{subequations} \label{4}
\begin{align}
        &G^\bot (q)\{\nabla_q \big(p^TM^{-1}(q)p\big)-M_dM^{-1}(q)\nabla_q \big(p^TM_d^{-1}(q)p\big)+2J_{2}M_d^{-1}p\}=0_{s\times 1},\label{4a}\\
        &G^\bot (q)\{\nabla_q V(q)-M_dM^{-1}\nabla_q V_d(q)\}=0_{s\times 1},\label{4b}
        %G^\bot (q)\{RM^{-1}p+(J_{20}-R_2)M_d^{-1}p\}=0,&
        \end{align}
    \end{subequations}
in which $s=n-m$ and $G^\bot$ is the left annihilator of $G$. Note
that, in the presence of physical damping in the model, there would
be one extra PDE, however, it will be removed with damping
injection~\cite{gomez2004physical}, by suitably selecting $K_v$. It
is to be noted that the analytical solution of the PDEs in \eqref{4}
requires extensive calculation, for which the detailed description
can be found in
\cite{donaire2016simultaneous,acosta2005interconnection}.
    Considering $H_d$ as the desired Lyapunov function of the closed loop system, its derivative will be:
    $$\dot{H}_d=-p^TM_d^{-1}R_2M_d^{-1}p,$$
    which is generally negative semi-definite. Thus the desired equilibrium point is stable. Equilibrium point is asymptotically stable if it is locally detectable from $G^TM_d^{-1}p$, in other words if $G^TM_d^{-1}p=0$. Note that if the following condition holds
    \begin{align}
    G^\bot(RM^{-1}M_d+M_dM^{-1}R)(G^\bot)^T>0,\label{5}
    \end{align}
    then $R_2$ is positive definite.

    %It is shown that one of the upper bounds of velocity
    %is related to the inverse of the smallest eigenvalue of $R_2$.
    %Note that the inequality (\ref{5}) is a
    %necessary and sufficient condition for the system (\ref{2}) to be
    %strongly dissipative \cite{gomez2004physical}, i.e. the damping matrix $R_2$ is positive definite.  The free parameters in (\ref{3})
    %is chosen such that boundedness of control law with a suitable bound
    %is ensured.
    % Due
    %to the difficulty of matching equations, in the proposed method
    %there is no need to find a solution with special properties for
    %$V_d$ and $M_d$. Thus, the computed upper bound for a specified system by IDA-PBC with different $H_d$ is different.
    %This is the expense  we shall pay to avoid resolving
    %PDEs (\ref{4}). We may improve the cases where the
    %upper bound of the control law is
    % very high by a two-phase controller (See Remark~\ref{re3} for a discussion
    %about this issue).

\section{Bounded Input IDA-PBC for Mechanical Systems}\label{s3}
In this section, having the parameters of the control law (\ref{3}),
we propose a methodology to find the upper bound of the control law.
First the upper bound of the potential energy terms is obtained by
suitably choosing the homogeneous solution of $V_d$. Next, to find
the upper bound of the kinetic energy, the upper bounds of the
velocity is found in Theorem \ref{th2}. Finally the upper bound of
the control law is obtained by adding up the upper bound of the
kinetic and potential energy related terms.

With a minor loss of generality, first assume that the elements of
the matrix $G$ are merely 0 or 1. The solution of the potential
energy PDE (\ref{4b}) may be separated by two terms as the
following:
    $$V_d=V_{dn}+V_{dh},$$
in which, $V_{dn}$ and $V_{dh}$ denote the
    non-homogeneous and homogeneous solution of the potential energy PDE,
    respectively. The homogeneous solution is given by:
    $$G^\bot M_dM^{-1}\nabla_q V_{dh}(q)=0,$$
    which may be represented as follows~\cite[ch. 2]{sneddon2006elements}
    \begin{equation*}
    V_{dh}=\Phi\big(V_{dh_{1}},V_{dh_{2}},\dots\big)
    \end{equation*}
where $\Phi$ is an arbitrary function and $V_{dh_{i}}$'s are the
solutions of the homogeneous equation. The function $\Phi$ usually
is chosen in quadratic form to satisfy (\ref{1.5}). However, this
may lead to unbounded control law. Thus, we propose an alternative
function for $\Phi$ in order to make the bounded control law.
Invoking~\cite{loria1997global}, this issue maybe is rectified by
defining $V_{dh}$ as follows
    \begin{equation}
    \label{6}
    V_{dh}=\displaystyle\sum k_i \int\mathcal{S}(V_{dh_{i}}-V_{dh_{i}}^*) d V_{dh_{i}}
    \end{equation}
    in which, $V_{dh_{i}}^*=\left.V_{dh_{i}}\right|_{q=q^*}$ and the
    function $\mathcal{S}(x)$ should satisfy the following properties
    \begin{itemize}
        \item[1.] $\mathcal{S}(0)=0$,
        \item[2.] $\mathcal{S}$ is an increasing function such that $|\mathcal{S}|\leq 1$,
        \item [3.] $\frac{d^2 \mathcal{S}(x)}{d x^2}\neq 0,
        \qquad \forall x\neq 0$.
    \end{itemize}
As an example, $S(x)=\tanh(x)$ satisfies these three conditions.
Generally, the control law of UMS might be unbounded, based on the
physics of the problem, as explained in Brockett's theorem (e.g.
when in cart-pole system the position of the swinging arm is
horizontal). In this situation regardless of the selection of $S(x)$
the functions $V_{dh}$ and     $V_{dn}$ could be non-smooth. Being
such, the following assumption is essential for development of the
rest of the paper (see Remark~\ref{re2} to see how this assumption
is relaxed).

    \begin{asu}\label{as1}
        \normalfont
        Consider the dynamic model of a mechanical system in the form (\ref{1}). Assume that the input coupling matrix is in the form $G=P[I_m,0_{s\times m}]^T$ with $P$ being a permutation matrix.
        The coefficients $k_i$s in \eqref{6} shall be sufficiently small. Considering the potential energy terms of the control law ($(\nabla_{q}V-M_dM^{-1}\nabla_q V_d)_i$), the upper
        bounds of the actuators should satisfy the following inequality:
        \begin{equation}\label{inequality}
        |(\nabla_{q}V-M_dM^{-1}\nabla_q V_d)_i|<\tau_{i_{max}}
        \end{equation}
        meaning that the control law should be able to compensate the effects of gravitational force in the entire workspace.
        In the above equation $(\cdot)_i$ denotes the $i$th element of $(\cdot)$ while
        $i$ counts merely for the actuated joints. In other words, $\nabla_q V$
        and $\nabla_q V_d$ are bounded as follow:
        \begin{equation}\label{potential}
        \|(\nabla_q
        V)_i\|\leq c_{V_{i}},\qquad\|\nabla_q V_d\|\leq c_{V_{d}}.
        \end{equation}
        \carrew
    \end{asu}
    Additionally for the kinetic energy terms of the control law ($\nabla_q K-M_dM^{-1}\nabla_q K_d$), as $K=1/2p^TM^{-1}p$ and $K_d=1/2p^TM_d^{-1}p$, it is clear that the kinetic energy terms are quadratic functions of $p$. Hence, restricting the velocity may result in the restriction of kinetic energy shaping terms in the control law. Before presenting the main results, we also need the following assumption.
    \begin{asu}\label{as2}\normalfont
        It is assumed that there are constants  $c_M$, $c_{M_{d}},c_J$ and
        $ c_{\Lambda_{i}}$ satisfying:
        \begin{equation}\label{assumption2}
        (\nabla_q K)_i\leq c_{M_{i}}\|p\|^2,\hspace{2mm} \nabla_q K_d\leq c_{M_{d}}\|p\|^2,\hspace{2mm} \|J_2(q,p)\|\leq c_J\|\tilde{p}\|,\hspace{2mm} \|\Lambda_i\|\leq c_{\Lambda_{i}}
        \end{equation}

        with $\Lambda:=M_dM^{-1}$.
        \carrew
    \end{asu}
    In the following theorem, the upper bounds of
    both $\|p\|$ and control law are found.
    Note that the previously mentioned assumption $G=P[I_m,0_{s\times m}]^T$ is relaxed in Remark~\ref{re1}.

    \begin{theorem}\label{th2} \normalfont
        Consider a mechanical system with the dynamic model (\ref{1}) and control law (\ref{3}). Presume that conditions (\ref{5}), Assumption~\ref{as1} and Assumption~\ref{as2} are satisfied, thus:
        %    Assume that with a minor loss of generality, $G=P[I_m,0_{s\times m}]^T$.
        \begin{itemize}
            \item [a)] The upper bound of $\|p\|$ and $\|\tilde{p}\|$
            are as follows:
            \begin{align}
            \label{norm}
            \|p\|\leq \sqrt{\frac{H_d(t_0)}{\lambda_m\{M_d^{-1}\}}}=
            c_{p1}, \quad \|\tilde{p}\|\leq \sqrt{\frac{H_d(t_0)}
                {\lambda_m\{M_d\}}}= c_{\tilde{p}1},
            \end{align}
            where $\tilde{p}=M_d^{-1}p$.
            \item[b)] The ultimate bound of $\|p\|$ and
            $\|\tilde{p}\|$ are given by:
            \begin{subequations}
                \begin{align}
                &   \|p\|\leq \Big(\dfrac{\lambda_M\{M_d^{-1}\}}{\lambda_m
                    \{M_d^{-1}\}}\Big)^{1/2}\frac{c_{V_{d}}\lambda_M
                    \{M_d^{-1}\}}{\lambda_m^2\{M_d^{-1}\}\lambda_m\{R_2\}+\mu}=
                c_{p2}.\label{9}
                \\
                &   \|\tilde{p}\|\leq \Big(\dfrac{\lambda_M\{M_d
                    \}}{\lambda_m\{M_d\}}\Big)^{1/2}\frac{c_{V_{d}}}{
                    \lambda_m\{R_2\}+\mu}= c_{\tilde{p}2}, \label{9.5}
                \end{align}
                in which $\lambda_m\{\cdot\}$ and $\lambda_M\{\cdot\}$ denote the maximum and the minimum eigenvalue of $\{\cdot\}$, respectively.
            \end{subequations}
            \item[c)] The upper bound of $|\tau_i|$ is:
            \begin{align}
            \label{13}
            c_{V_{i}}+c_{\Lambda_{i}}c_{V_{d}}+\big(c_{M_{i}}+c_{\Lambda_{i}} c_{M_{d}}\big)c_p^2+c_Jc_{\tilde{p}}^2+\lambda_M\{K_v\}c_{\tilde{p}},
            \end{align}
            where $c_p$ and $c_{\tilde{p}}$ denote the upper bound of $\|p\|$ and $\|\tilde{p}\|$ respectively.
            Generally, $c_p=c_{p1}$ and $c_{\tilde{p}}=c_{\tilde{p}1}$, but if $\|p(t_0)\|\leq c_{p2}$ and $\|\tilde{p}(t_0)\|\leq c_{\tilde{p}2}$, then
            $$c_p=\min \{c_{p1},c_{p2}\}, \qquad c_{\tilde{p}}=\min \{ c_{\tilde{p}1}, c_{\tilde{p}2}\}.$$
            \carrew
        \end{itemize}
    \end{theorem}
    \textbf{\textit{Proof}}. a) Consider $H_d$ as a Lyapunov candidate. Since
    the time derivative of $H_d$ is negative semi-definite, the system's trajectory that
    starts from $H_d(t_0)$  descends to lower level sets. In other words, we
    have the following inequality.
    $$\frac{1}{2}\lambda_m\{M_d\}\|\tilde{p}\|^2\hspace{2mm}\&\hspace{2mm}\frac{1}{2}\lambda_m\{M_d^{-1}\}\|p\|^2\leq K_d\leq H_d\leq H_d(t_0).$$
    Hence, the upper bounds of $|\tilde{p}\|$ and $\|p\|$ are derived easily.

    b) Consider the desired kinetic energy as a Lyapunov candidate:
    \begin{equation}
    \label{7}
    K_d=H_d-V_d,
    \end{equation}
    and its time derivative as:
    \begin{align}
    \label{8} \dot{K}_d&=-p^TM_d^{-1}R_2M_d^{-1}p-(\nabla_q
    V_d)^TM_d^{-1}p\nonumber\\&\leq
    -\|p\|^2\lambda_m^2\{M_d^{-1}\}\lambda_m\{R_2\}+\|\nabla_q
    V_d\|\lambda_M\{M_d^{-1}\}\|p\|.
    \end{align}
    This inequality assures that $\|p\|$ is bounded since the negative
    part is proportional to $\|p\|^2$ while the other term is
    proportional to $\|p\|$. In order to derive the upper bound of
    $\|p\|$, Theorem 4.18 of~\cite{khalil2002nonlinear} is utilized. It
    is clear that Lyapunov candidate (\ref{7}) is within the following
    bounds
    $$\frac{1}{2}\lambda_m\{M_d^{-1}\}\|p\|^2\leq K_d\leq \frac{1}{2}\lambda_M\{M_d^{-1}\}\|p\|^2.$$
    On the other hands, the right hand side of (\ref{8}) is less than $-\mu\|p\|^2$ if:
    $$\|p\|\geq \frac{c_{V_{d}}\lambda_M\{M_d^{-1}\}}{\lambda_m^2\{M_d^{-1}\}\lambda_m\{R_2\}+\mu},$$
    where $\mu\in\mathbb{R}^+$ is an arbitrary value, while
    Assumption~\ref{as1} is used for the upper bound of $\nabla_q V_d$. The
    final ultimate bound for $\|p\|$ is derived as (\ref{9}). With a
    similar approach, the ultimate bound of $\tilde{p}$ can be found
    . Note that it is possible to compute upper bound of
    $\tilde{p}$ based on (\ref{9}), however it is a more conservative
    choice than (\ref{9.5}). From (\ref{9}) and (\ref{9.5}), it is clear that the
    ultimate bound of velocity can be reduced by increasing
    $\lambda_m\{R_2\}$. Notice that these ultimate bounds are also the
    upper bounds if the initial condition of $p$ and $\tilde{p}$ are less
    than $c_{p2}$ and $c_{\tilde{p}2}$, respectively.

    c) Based on Assumption~\ref{as2}, the
    upper bound for $i$th element of the kinetic energy shaping terms is:
    \begin{equation}
    \label{10}
    \big(c_{M_{i}}+c_{\Lambda_{i}} c_{M_{d}}\big)c_p^2%+\eta\gamma^2)+\\\lambda_M\{K_v\}\lambda_M\{M_d^{-1}\}\gamma\triangleq \varepsilon+\lambda_M\{K_v\}\lambda_M\{M_d^{-1}\}\gamma
    \end{equation}
    in which the parameters was defined in the Assumption~\ref{as2}. Note that
    as explained in~\cite{acosta2005interconnection}, $J_2$ should be
    linear with respect to velocity. Hence, it can be represented as
    $$J_2(q,p)=\displaystyle\sum_{i=1}^{n_0}\tilde{p}^T\alpha_i(q)W_i$$
    where $n_0=\frac{n(n-1)}{2}$, and $W_i$s are skew-symmetric constant
    matrices. Since $\alpha_i$s are determined based on PDE (\ref{4a}),
    we infer that:
    \begin{equation}\label{J2}
    \|J_2(q,p)\|\leq c_J\|\tilde{p}\|
    \end{equation}

    Thus,
    $J_2\tilde{p}$ has also an upper bound proportional to
    $\|\tilde{p}\|^2$. Now the damping term in (\ref{3}) is analyzed for which the upper bound is given by:
    \begin{align}
    \label{12}
    \|K_vG^T\tilde{p}\|\leq \lambda_M\{K_v\}c_{\tilde{p}}.%=\Big(\dfrac{\lambda_M\{M_d\}}{\lambda_m\{M_d\}}\Big)^{1/2}\frac{c_{V_{d}}\lambda_M\{K_v\}}{\lambda_m\{R_2\}+\mu}
    \end{align}
    Finally, the upper bound of $i$th element of control law (\ref{3})
    is derived simply by adding the above terms obtained in \eqref{12},\eqref{10},\eqref{J2},\eqref{assumption2} and \eqref{potential} as (\ref{13}).
    \carre

    \begin{remark}\normalfont
The natural damping of the system represented by $R$ is
configuration independent, whether it is actuated or not, thus $R$
may be considered positive definite. If the natural damping is
considered in the modeling, condition (\ref{5}) is certainly
satisfied. On the other hand, if it is not modeled such, or if
(\ref{5}) is not    satisfied, part $b$ of the Theorem~\ref{th2} is
not applicable. As explained earlier, the proposed method finds two
upper bounds for the velocity. For the model with no natural damping
term, the upper bound of the velocity is obtained from part $a$ of
the Theorem~\ref{th2}. On the other hand, if the natural damping is
included in the model and $R_2$ is positive definite, the bound of
the velocity will be the minimum of the two upper bounds obtained in
parts $a$ and $b$ of the Theorem~\ref{th2}.
        \carrew
\end{remark}

    In some UMS like the parallel robot in~\cite{harandi2017motion} and VTOL
    aircraft~\cite{acosta2005interconnection}, the input coupling matrix $G(q)$ is
    configuration-dependent. Furthermore, the bound of the control input is not
    necessarily symmetric ($\tau_{min}\neq-\tau_{max}$), e.g., in cable-driven
    robots, the control input $\tau$ should be positive as the cables merely
    apply tensile forces~\cite{harandi2019stabilization}. In the
    following remark, these issues are addressed.

    \begin{remark}\label{re1}\normalfont
        Let $G(q)$ be a full rank matrix that maps the actuator
        forces/torques to configuration space force/torque of the robot. Hence,
        without  loss of generality, as it is configuration dependant,
        it is bounded in the entire workspace of the
        robot. Assume that,
        $$ \|(G^TG)^{-1}G^T\|\leq G_M,\qquad    \|G\|\leq G_m.$$
        In this case, the upper bound for the control law will be given by
        \begin{align}
        G_M\big(c_{V_{i}}+c_{\Lambda_{i}}c_{V_{d}}+\big(c_M+c_{\Lambda_{i}}
        c_{M_{d}}\big)c_p^2+c_Jc_{\tilde{p}}^2\big)&\nonumber\\+ G_m\lambda_M\{K_v\}c_{\tilde{p}},&
        \label{g}
        \end{align}
        which is more conservative than (\ref{13}). For the case of non-symmetric bounds on the control input, the lower bound of
        the actuators should be calculated separately. In this case, assume that $\sigma$ is the minimum of the following
        expression, in the
        workspace of the robot,
        \begin{equation*}
        {\sigma}\leq (G^TG)^{-1}G^T\bigg(\nabla_q V-M_dM^{-1}\nabla_q V_{d}\bigg).
        \end{equation*}
        Hence, using \eqref{g} the lower bound of the control law \eqref{3} is,
        \begin{equation*}
        {\sigma}-G_M\big((c_{M_{i}}+c_{\Lambda_{i}} c_{M_{d}})c_p^2+c_Jc_{\tilde{p}}^2\big)-G_m\lambda_M\{K_v\}c_{\tilde{p}}.
        \end{equation*}
        Thus, it is proved that, for the cable robot, if the above-mentioned lower bound is positive, it is sufficient to ensure the boundedness of the lower bound. It should also be noted that for some UMS, one may derive an upper bound which is less
        conservative than what is given in (\ref{g}) (see the example in section~\ref{s42}). \carrew
    \end{remark}
    One of the main features of the IDA-PBC is that if the output is
    detectable, asymptotic stability is guaranteed~\cite{donaire2016simultaneous}. However, the asymptotic stability may require high control
    effort in some conditions. This fact is in confirmation of the Brockett's
    theorem~\cite{brockett1983asymptotic} which expresses that for the UMS,
    the control law could be non-smooth, such as the sace of cart-pole represented in~\cite{spong1996energy}.
    It is due to the fact that when the states of the system
    are close to a singular region, the control effort will get large values.
    As an example, for the cart-pole
    system~\cite{acosta2005interconnection},
    when the pendulum is close to the horizontal configuration,
    the control effort should get very large values to prevent
    the pendulum from falling. The following remarks are given on
    this issue.
    \begin{remark}\label{re2}\normalfont
        In the Assumption~\ref{as1}, $\nabla_q
        V_d$ is presumed to be bounded. However, in some UMS, $\nabla_q
        V_d$ is non-smooth. Letting $H_d$ to be the Lyapunov function, the
        system trajectory starts on the level set $H_{d}(t_0)$ and goes to
        lower level sets until reaching zero. The maximum value of $q_i-q_{i}^*$,
        may be found by setting $p=0$ and $q_j=q_{j}^*$ for
        $j=1,...,i-1,i+1,...,n$ in $H_d$ and set it equal to $H_{d}(t_0)$. By
        this means, the upper bound for each elements of $\nabla_q V_d$ is
        derived. Hence,  Assumption~\ref{as1} is always satisfied. This has been used in the VTOL example explained in \ref{s42}
        %\carrew
    \end{remark}
    \begin{remark}\label{re3}\normalfont
        The maximum torque/force of actuators is practically limited, and it
        might be lower than the required upper bound proposed in (\ref{13}), causing actuator saturation.
        This limitation on the control input adversely affects the performance and the stability of the closed-loop system. In other words, this may lead to a divergence of the
        configuration variables from the region of attraction. Based on
        Remark~\ref{re2} large control efforts might be due to an initial configuration of the system in the vicinity of the singular regions. To
        remedy this problem, we propose to design a two-phase controller,
        such that the first phase aims to bring the system into a configuration
        away from the singular point (\ref{13}) and then
        IDA-PBC will be taking over the control. The design of the first
        controller is case-dependent and might be done by prior knowledge of
        the system, but usually not a prohibitive task.
    \end{remark}

    \subsection{Discussion}
In this section, we discuss how the obtained bounds of the control
input can be used to design a suitable controller in order to keep
the control effort within the actuator limit. Additionally, to
validate the proposed method, a comparison is made to similar
studies in the literature, in which the upper bounds of the control
input are found for specific case studies. Furthermore, we explain
the applicability of the proposed method for fully actuated systems.

In \eqref{13}, except the upper bound of the gravity ($c_{V}$), all
the other parameters in \eqref{13} are based on the solution of the
PDEs in (\ref{4}) and the free design gains of the controller, $K_v$
and $k_i$s defined in \eqref{6}. Hence, the selection of the
suitable solution of the PDEs and gains of the controller gives the
possibility to design the controller considering the limitation of
the control input. The smaller values for $k_i$s result to lower
control input and its bound, with the expense of increasing the
convergence time. On the other hand, the effects of $K_v$ can be
considered two folds, by increasing $K_v$ it is expected that the
control law will also increase. However, it may reduces the upper
bound of the velocity; thus increasing $K_v$ can also decrease the
upper bound of the control input. To better investigate the effect
of $K_v$, we first discuss its effect in the fully actuated systems,
then we classify UMS into two categories and discuss each case in
detail. Our proposed methodology may also be extended for the fully
actuated systems without having to solve the PDEs and having the
condition (\ref{5}). Note that in most  cases $c_{p1}<c_{p2}$ and it
may be inferred that $c_{p2}$ is not  required, but in total energy
shaping of fully actuated robots, we can have a smaller value for
$c_{p2}$ by assigning larger values for $K_v$. It is due to the fact
that in a fully actuated system, by increasing $K_v$, the
$\lambda_m\{R_2\}$ can be increased, while in the UMS, this is not
necessarily valid. Thus, with $K_v=\kappa I_n$ and considering a
large value for $\kappa$, the upper bound of the fully actuated
systems is approximated by the following expression:
    $$
    c_{V_{i}}+c_{\Lambda_{i}}c_{V_{d}}+c_{V_{d}}\Big(\dfrac{\lambda_M\{M_d\}}{\lambda_m\{M_d\}}\Big)^{1/2}.$$

for the case UMS, $K_v$ may be chosen based
    on the following discussion.
    If $ c_{\tilde{p}}= c_{\tilde{p}2}$,  the right hand side of
    (\ref{12}) is equal to,
    \begin{equation}\label{fraction}
    \lambda_M\{K_v\}c_{\tilde{p}}=\Big(\dfrac{\lambda_M\{M_d\}}
    {\lambda_m\{M_d\}}\Big)^{1/2}\frac{c_{V_{d}}\lambda_M\{K_v\}}{\lambda_m\{R_2\}+\mu}
    \end{equation}

By considering $\mu$ to be infinitesimal ($\mu\approx 0$), and
replacing $R_2$ using \eqref{4.5} we have the following expression:
    \begin{align*}
    \frac{\lambda_M\{K_v\}}{\lambda_m\{R_2\}} =\frac{\lambda_M\{K_v\}}{\lambda_m\{\frac{1}{2}(RM^{-1}M_d+M_dM^{-1}R)+GK_vG^T\}}
    \end{align*}
Thus, to have a smaller upper bound for the damping injection term
of the control law, if ($RM^{-1}M_d+M_dM^{-1}R$) is positive
definite, $K_v$ shall be as small as possible. On the other hand, if
($RM^{-1}M_d+M_dM^{-1}R$) is not positive definite, then $K_v$ shall
be sufficiently large to results in $R_2>0$. From (\ref{9}),   it is
also inferred that the upper bound of the velocity is reduced by
having a larger value for $\lambda_m\{R_2\}$. In this situation if
($RM^{-1}M_d+M_dM^{-1}R\geq0$) and ${\lambda_m\{R_2\}}$ is a
function of $K_v$ analytically finding a suitable $K_v$  is not
straight forward, and it shall be found case by case. However, with
$GK_vG^T$ being positive semi-definite if
$(RM^{-1}M_d+M_dM^{-1}R)>0,$  and $\lambda_m\{R_2\}$ is not affected
by $K_vG^T\tilde{p}$, then $K_v=0$ is the best choice for minimizing
the upper bound (\ref{13}). Generally, we can argue that     by
calculating the limit of \eqref{fraction} for $K_v\to 0$ and $K_v\to
\infty$, then it is possible to assign a value for $K_v$ such that,
\eqref{fraction} will be smaller than 1.

For the comparison of the obtained method to the results reported in
the literature, it is to be noted that the upper bound obtained by
(\ref{13}) is the generalization of the methods for the upper bound
of fully actuated serial
robots~\cite{loria1997global,zavala2007natural} and the class of
underactuated robots considered in~\cite{ortega2013passivity}. By
merely considering the potential energy shaping and having a
sufficiently small value for $K_v$ using (\ref{13}) gives the same
results as
in~\cite{loria1997global,zavala2007natural,ortega2013passivity}.
Additionally, in the total energy shaping for a special case of
underactuated robots with constant $M$ and $M_d$, the upper bound
(\ref{13}) of our proposed method is simplified as,
$$c_{V_{i}}+c_{\Lambda_{i}}c_{V_{d}}+\lambda_M\{K_v\}c_{\tilde{p}},$$
which is the same results as obtained in \cite{santibanez2005control}.

\section{Simulation Results}\label{s4}
In this section, application of the proposed method is evaluated by
applying it to two UMS. The first case study is the well-known ball
and beam benchmark system, in which the position of the ball on the
beam and the angle of the beam are controlled with a given input to
the beam angle. The second case study is VTOL which has three DOF
with two control inputs. In this case study, we consider two
solutions of the PDEs and compare the upper bound obtained via each
of the solutions. Additionally, we propose a two-phase controller to
deal with non-smooth control law.

\subsection{Ball and Beam System}
\begin{figure}[t]
        %{\includegraphics[width=9cm]{a2}}
        \includegraphics[width=.99\linewidth]{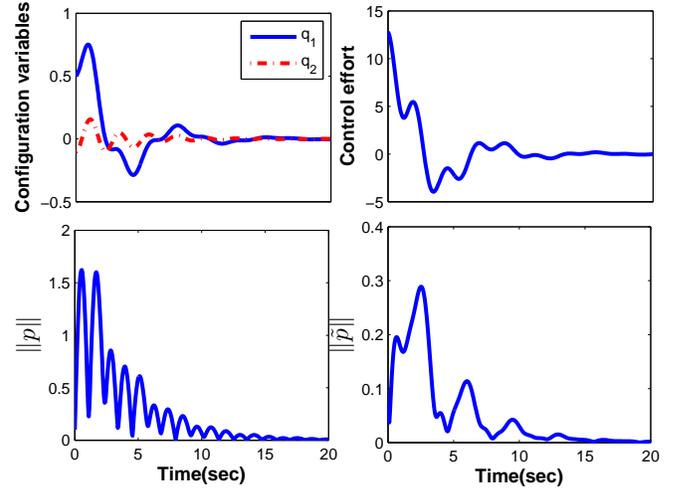}
        \centering
        \caption{Simulation results of the ball and beam system.
        } \label{p1}
    \end{figure}

The system consists of a ball moving along a beam whose angle is
controlled.  The position of the ball with respect to the pivot
point of the beam is denoted by  $q_1$, and the angle of the beam
with respect to the vertical line is given by $q_2$. The dynamic
parameters of the system are chosen based on the model given
in~\cite{ortega2002stabilization},
    $$M=\begin{bmatrix}
    1 & 0 \\ 0 & L^2+q_1^2
    \end{bmatrix},\qquad\quad V=gq_1\sin(\theta),$$
    where $L$ is the half length of the beam, $\theta$ is the beam
    angle, and $g$ is the gravity constant. For the control law we utilize the proposed controller
    in~\cite{ortega2002stabilization} with the desired parameters of the closed-loop system, given by:
    \begin{align*}
    M_d&=\begin{bmatrix}
    \sqrt{2}(L^2+q_1^2)^{1/2} & L^2+q_1^2 \\ L^2+q_1^2 & \sqrt{2}(L^2+q_1^2)^{3/2}
    \end{bmatrix}\\
    V_d&=g[1-\cos(q_2)]+\frac{k_p}{2}\big[q_2-\frac{1}{\sqrt{2}}\mbox{arcsinh}(\frac{q_1}{L})\big]^2\\
    J_2&=\begin{bmatrix}
    0 & j \\ -j & 0
    \end{bmatrix},\quad j=q_1[p_1-\sqrt{2}(L^2+q_1^2)^{-1/2}p_2],
    \end{align*}

in which $L=2m$, $R=\mbox{diag}\{0.2,0.1\}$, and the gains of the
controller are $k_p=5$ and $k_v=5$ based on
\cite{gomez2004physical}. The initial condition of the robot is
considered to be $[q^T(0),p^T(0)]=$ $[0.5,-0.1,0.1,0]$ based on
\textit{Proposition 5} of \cite{ortega2002stabilization} and the
controller is supposed to position the ball at the equilibrium point
$[q^T(0),p^T(0)]=$ $[0,0,0,0]$. With the given parameters,
$H_d(0)=0.24$ is less than $0.31$ derived from \textit{Proposition
5} of  \cite{ortega2002stabilization}, which ensures that the ball
is always on  the beam. Note that we do not modify the homogeneous
solution of $V_d$ to utilize the analysis proposed in
\cite{ortega2002stabilization}. The parameters in Theorem~\ref{th2}
are derived after some cumbersome calculation as follows:
\begin{align*}
    &c_{V_{2}}=10.4,\qquad\quad
    c_{V_{d}}=2.4,\qquad\quad
    c_{\gamma_{2}}=6,\\  &c_{M_{2}}=0,\hspace{2mm}
    \quad\qquad c_{M_{d}}=0.9, \qquad\quad
    c_J=10.4,\\ &\lambda_M\{M_d^{-1}\}= 0.82, \qquad\quad
    \hspace{1mm} \lambda_m\{M_d^{-1}\}=0.06
    \end{align*}
Using \eqref{norm} one  can simply compute $c_p=2$ and
$c_{\tilde{p}}=0.44$, thus the upper bound of the control input will
be $20$. The   simulation results are depicted in Fig.~\ref{p1}. The
control effort is less than 15 which is clearly less that than the
upper bound obtained from
    (\ref{13}). The upper bound of
    $\|p\|$ and $\|\tilde{p}\|$ are about 1.6 and 0.3 which are less
    than $c_P$ and $c_{\tilde{p}}$, respectively. Notice that the reason
    of difference between the actual and the computed upper bound is that in
    (\ref{13}) we consider the worst-case scenario in obtaining the upper bounds.
    \subsection{VTOL Aircraft}\label{s42}
    In the previous example, the effects of velocity-dependent  terms
    was investigated. Through this example we aim to analyze the Remarks~\ref{re1},~\ref{re2}  and~\ref{re3}.
    Hence, we consider an UMS with strongly coupled dynamics, referred to as VTOL. First a non-smooth control law is derived and the tracking performance of the system and the magnitude of the control effort is analysed. Secondly we propose a two-phase controller to compensate for the effects of non-smooth control law. Finally for the comparison purpose the performance of a smooth control law is also investigated.  The dynamic model of the VTOL is given in \cite{acosta2005interconnection} as follow:

    \begin{equation*}
    %\label{26}
    M=I, \quad
    V=gy,\quad  q=\begin{bmatrix}
    x\\y\\ \theta
    \end{bmatrix},\quad
    G(q)=
    \begin{bmatrix}
    -\sin(\theta) & \epsilon\cos(\theta) \\
    \cos(\theta) & \epsilon\sin(\theta) \\
    0 & 1
    \end{bmatrix}
    \end{equation*}
    where $x,y$ is the position of the center of gravity,  $\theta$ is the
    roll angle and $\epsilon$ denotes the effect of the slopped wings.
    The purpose is to stabilize the unstable equilibrium point
    $[x^*,y^*,0]$ with bounded input IDA-PBC controller. In
    \cite{acosta2005interconnection} an IDA-PBC controller with
state-dependent $M_d$ and smooth $V_d$ is designed. Here, in order
to analyze the effects of non-smooth terms, intentionally a
    locally stabilizing controller with constant $M_d$ and non-smooth
    $\nabla_qV_d$ is applied. The parameters of the controller are chosen as:
    \begin{align*}
    M_d&=\begin{bmatrix}
    20\epsilon^2 & 0 & \epsilon \\ 0 & 1 & 0 \\ \epsilon & 0 & 0.1
    \end{bmatrix},\qquad
    J_2=0_{3\times 3},\\
    V_d&=k_1\ln\cosh\Big(\epsilon(y-y^*)+\ln\big(\epsilon\cos(\theta)-0.1\epsilon\big)\Big)\\&+k_2\ln\cosh\Big(\frac{1}{20\epsilon}(x-x^*)-\theta-0.1\text{arctanh}\big(1.1055\tan(\frac{\theta}{2})\big)\Big)\\&-k_1\epsilon\tanh\ln(0.9\epsilon)(y-y^*)\\&-\frac{g+k_1\epsilon\tanh\ln(0.9\epsilon)}{\epsilon}\ln\big(\epsilon\cos(\theta)-0.1\epsilon\big)-\rho,
    \end{align*}
    with $k_1,k_2\in\mathbb{R}^+$ being the free parameters and the constant $\rho$ is determined such that $V_d(t_0)=0$.
    It is clear that the terms $\text{arctanh}\big(1.1055\tan(\frac{\theta}{2})\big)$ and $\ln\big(\epsilon\cos(\theta)-0.1\epsilon\big)$ confine $\theta$ in a region inside the workspace and lead to non-smooth terms in $\nabla_q V_d$. Hence, we invoke the results of Remark~\ref{re2} and \ref{re3}.
    Note that due to simple structure of $G$, it is possible to derive better upper bounds with respect to the conservative upper bound proposed in (\ref{g}) by the following matrices,
    \begin{align*}
    (G^TG)^{-1}G^T&=\begin{bmatrix}
    -\sin(\theta) & \cos(\theta) & 0 \\
    \frac{\epsilon\cos(\theta)}{1+\epsilon} & \frac{\epsilon\sin(\theta)}{1+\epsilon} & \frac{1}{1+\epsilon}
    \end{bmatrix}\\
    (G^TG)^{-1}G^TM_dM^{-1}&=\begin{bmatrix}
    -20\epsilon^2\sin(\theta) & \cos(\theta) & -\epsilon\sin(\theta) \\
    \frac{20\epsilon^3\cos(\theta)+\epsilon}{1+\epsilon} & \frac{\epsilon\sin(\theta)}{1+\epsilon} & \frac{\epsilon^2\cos(\theta)+0.1}{1+\epsilon}
    \end{bmatrix}
    \end{align*}
    \begin{figure}[!t]
        %{\includegraphics[width=9cm]{a2}}
        \includegraphics[width=.9\linewidth]{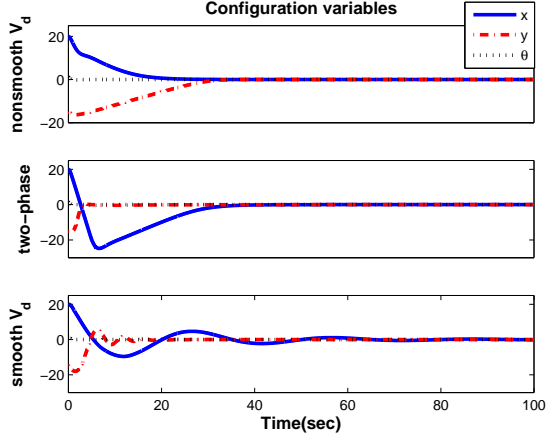}
        \centering
        \caption{Configuration variables of the VTOL aircraft.
        } \label{p2}
    \end{figure}
    \begin{figure}[!b]
        %{\includegraphics[width=9cm]{a2}}
        \includegraphics[width=.9\linewidth]{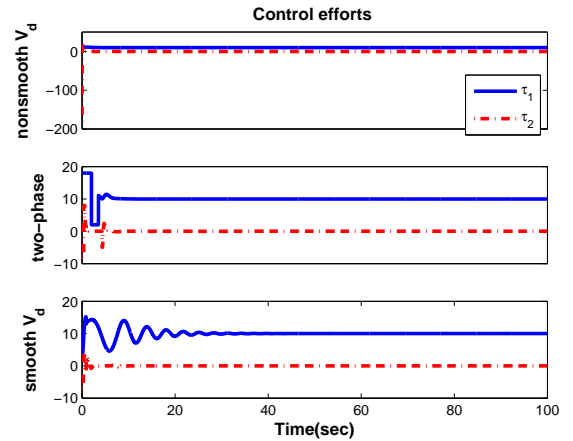}
        \centering
        \caption{Control efforts of the VTOL aircraft.
        } \label{p3}
    \end{figure}
    In other words, a constant gravity is applied to the $y$ direction
of the system. Thus, the bounds of controller are in the following
form:
    \begin{align}
    |\tau_1-g|&\leq \max_q\{|g-\big((G^TG)^{-1}G^T\nabla V\big)_1|\} \nonumber\\&+\max_q\{ \|(G^TG)^{-1}G^TM_dM^{-1}\|\}c_{V_d}+\lambda_M\{K_v\},\nonumber\\
    |\tau_2|&\leq \max_q\{|\big((G^TG)^{-1}G^T\nabla V\big)_2|\} \nonumber\\&+\max_q\{ \|(G^TG)^{-1}G^TM_dM^{-1}\|\}c_{V_d}+\lambda_M\{K_v\},
    \label{ub}
    \end{align}
where the damping term is modified to be
$K_v\mathcal{S}(G^T\tilde{p})$ with $\mathcal{S}$ defined in
(\ref{6}).
    The initial condition of the system is $[20,-15,1.3]^T$ and the desired position is the equilibrium point of the system at $[0,0,0]^T$.
    Since the initial value of $\theta$ is close to singularity, it results into high values for the control effort. Invoking Remark~\ref{re2} one can compute the upper bound of $\theta=1.33$ by considering $k_1=4$ and $k_2=5$. By this means, the following upper bounds are derived after some calculations,
    \begin{align*}
    &|\big((G^TG)^{-1}G^T\nabla V\big)_1|\leq 10,\quad |\big((G^TG)^{-1}G^T\nabla V\big)_2|\leq 2.25,\\
    &\|(G^TG)^{-1}G^TM_dM^{-1}\|\leq 1.75,\quad c_{V_{d}}=170
    \end{align*}
    %\begin{figure}[t]
    %{\includegraphics[width=9cm]{a2}}
    %  \includegraphics[width=.99\linewidth]{f4.eps}
    % \centering
    %\caption{Simulation results of the VTOL aircraft.
    %} \label{p}
    %\end{figure}
    Simulation results with {$K_v=I_2$} and $\mathcal{S}=\tanh$ is
    depicted in Fig.~\ref{p2} and Fig.~\ref{p3}, denoted by `{{nonsmooth $V_d$}}' in both figures. The simulation results of the two-phase controller and the smooth control law are also demonstrated in figures identified with the label `{{two-phase}}' and `{{smooth $V_d$}}', respectively. In Fig.~\ref{p2} the tracking performance of the controller is shown, and Fig.~\ref{p3} shows the control effort. 
As it is clear from the figures, for the non-smooth control law, the configuration variables converge to zero but the upper bound of the control effort is about 200 which is not
even close to being applicable. To rectify this problem, a two-phase
controller based on Remark~\ref{re3} is designed. Additionally,  the
controller proposed in \cite{acosta2005interconnection} is
simulated. In the two-phase controller,
    the primary controller is used in the first phase to converge $\theta$ to zero. Hence, for a small value of the
    $\theta$, the dynamic of the system can be represented in the following form:
    $$\ddot{x}\approx    0,\qquad\qquad \ddot{y}\approx \tau_1-g,$$
    which is obtained by replacing $\theta\approx0,\tau_2\approx   0$ in the dynamic of the system. Considering the simplified dynamic, then a secondary controller is designed using IDA-PBC approach such that $y-y^*$ converges
    to zero. Although with this controller $|x|$ will increase with a
    constant speed, it is possible to reduce $H_d(t_0)$ arbitrary by
    increasing $M_{d_{(1,1)}}$ or decreasing $k_2$ in which $t_0$ denotes the
    initial time of applying the secondary controller. In the two-phase controller the primary controller is given by:
    \begin{align*}
    \tau_1&=g-\varkappa_1\mathcal{S}(\kappa_1 y+\kappa_2 \dot{y})\\
    \tau_2&=-\varkappa_2\mathcal{S}(\kappa_3 \theta+\kappa_4 \dot{\theta})
    \end{align*}
    with the following parameters:
    \begin{align*}
    \varkappa_1=\varkappa_2=8,\hspace{2mm} \kappa_1=30,\hspace{2mm}\kappa_2=20,\hspace{2mm}\kappa_3=80,\hspace{2mm}\kappa_4=30,
    \end{align*}
    with this aim that $|\tau_1-g|\leq 10$ and $|\tau_2|<10$.
    Referring to the simulation results in Fig.~\ref{p2} and Fig.~\ref{p3}, it is clear that after applying the IDA-PBC, the control
    efforts are in a the bounds, and configuration variables converge
    to their desired values. This shows the superiority of this method.
    Note that one may design a better controller for the first phase to
    prevent the increase of $|x|$, but this is out of the scope of
    this paper. We have also investigated the controller proposed in \cite{acosta2005interconnection} which is based on a smooth $V_d$. The parameters of the controller are $P=\mbox{diag}\{0.003,0.005\}$ and $K_v=0.2I_2$ . The simulation results shows that $|\tau_1-g|$ and $|\tau_2|$ lies in the predefined range. It should be noted that, in this case we have considered smaller gains for the controller to have smaller control effort, in the expense of slow convergence rate. This study confirms that the suitable solution for the PDEs is an essential part of the controller design while considering bounded input control problems.

    \section{Conclusion}
In this paper we addressed the bounded input control of mechanical
systems based on IDA-PBC approach. In IDA-PBC, the control law is
formulated in terms of the the desired potential and kinetic energy
for the closed-loop system, and we found the upper bound of each
component. The desired potential and kinetic energy of the IDA-PBC
are the solution of PDE and the upper bound of the potential energy
are found by suitably defining the homogeneous solution of desired
potential energy. For the upper bounds of the kinetic energy related
term, the velocity related terms are complex to deal with and we
proposed a method to find the upper bounds of the velocity. It is
also discussed that how the gains of the controller and the solution
of the PDEs are to be chosen and how the selection of the design
parameters affects the performance of the closed-loop system. The
bounded input control of IDA-PBC control is investigated for
potential energy shaping as well as constant mass matrix systems in
literature, and the proposed method includes kinetic energy shaping
for configuration dependant mass matrix. The proposed method is also
extended for systems with non-symmetric control bounds as well as
for the systems with state-dependent input coupling matrix.
Moreover, we proposed a methodology to deal with non-smooth control
inputs with a two-phase control scheme. The validity and the
application of the proposed method is further investigated through
simulation studies for two benchmark underactuated mechanical
systems.

%    \section*{Disclosure Statement}
%No potential conflict of interest was reported by the authors.
%    \section*{Funding}
%This work was supported by the Iranian National Science Foundation
%(INSF) under Grant No. 92006640.

    \bibliography{ref}
\bibliographystyle{IEEEtran}
\end{document}